# Precision modeling of M dwarf stars: the magnetic components of CM Draconis

J. MacDonald and D. J. Mullan

Dept of Physics and Astronomy, University of Delaware, Newark DE 19716, USA

**Abstract**

The eclipsing binary CM Draconis contains two nearly identical red dwarfs of spectral class dM4.5. The masses and radii of the two components have been reported with unprecedentedly small statistical errors: for *M*, these errors are 1 part in 260, while for *R*, the errors reported by Morales et al (2009) are 1 part in 130. When compared with standard stellar models with the appropriate masses and age (≈4 Gyr), the empirical results indicate that, when compared to standard stellar models, both components are larger in *R*, and lower in luminosity *L*, by several standard deviations. Here, we attempt at first to model the two components of CM Dra in the context of standard (non-magnetic) stellar models using different assumptions about helium abundances (*Y*), heavy element abundances (*Z*), opacities, and mixing length parameter (*α*). We find no 4 Gyr-old models with plausible values of these 4 parameters which fit *L* and *R* within the reported statistical error bars.

However, CM Dra is known to contain magnetic fields, as evidenced by the occurrence of starspots and flares. Here we ask: can inclusion of magnetic effects into stellar evolution lead to fits of *L* and *R* within the error bars? Morales et al. (2010) have reported that the presence of polar spots results in a systematic overestimate of *R* by a few percent when eclipses are interpreted with a standard code. In a star where spots cover a fraction *f* of the surface area, we find that the revised *R* and *L* for CM Dra A can be fitted within the error bars by varying the parameter *α*. The latter is often assumed to be reduced by the presence of magnetic fields, although the reduction in *α* as a function of *B* is difficult to quantify. An alternative magnetic effect, namely, inhibition of the onset of convection, *can* be readily quantified in terms of a magnetic parameter $\delta \approx B^2/4\pi\gamma p_{gas}$. In the context of *δ* models in which the field strength is not allowed to exceed a "ceiling" of $10^6$ gauss, we find that the revised *R* and *L* can also be fitted, within the error bars, in a finite region of the *f* – *δ* plane. The permitted values of *δ* near the surface lead us to estimate that the vertical field strength on the surface of CM Dra A is about 500 Gauss: the total field strength on the surface will exceed this.

## 1. Introduction

As stellar parameters are being derived with ever increasing precision, the task of fitting the observations with stellar models is becoming ever more challenging. Examples of this are provided by ongoing observations of eclipsing binaries: these can lead to radii and masses for many stars being extracted with unprecedented precision. Torres et al. (2010) report on 95 systems in which the masses and radii of both stars in the binary are known to ±3% or better. Comparison between empirical data and standard stellar models indicates that, for some low mass stars, the empirical radii are larger than the models predict by ~5-10% , and the empirical effective temperatures are smaller than the models predict by ~3-5% (e.g. Morales et al. 2010).

Because of the precision of the data, the discrepancies in empirical radii among low-mass stars are by no means insignificant effects: on the contrary, the discrepancies in radii may amount to many standard deviations relative to the standard models. A case in point is CM Draconis, an eclipsing binary containing two dM4.5 stars with masses of $0.23102\pm0.00089\,M_\odot$ and $0.21409\pm 0.00083\,M_\odot$, and empirical radii of $0.2534\pm0.0019\,R_\odot$ and $0.2398\pm0.0018\,R_\odot$ respectively (Morales et al. 2009; Torres et al. 2010). Plotting these radii against mass, and comparing with stellar models having a range of ages and heavy element abundances, it emerges that both components have radii which are larger than main sequence models predict by at least $0.01\,R_\odot$ (Morales et al. 2009). In view of the small statistical errors which are quoted for the empirical radii (i.e. 1 part in 133 for CM Dra A, and 1 part in 133 also for CM Dra B), the "bloating" of the radii of both components is at least a 5σ effect.

Since the components of CM Dra have the smallest known accurate masses for main sequence M-dwarf stars, this is an important and challenging system for precision testing of stellar evolutionary models on the lowest regions of the main sequence. In order to set the stage for our modeling efforts, we first summarize some relevant observational properties of CM Dra.

### 1.1. Observational properties of CM Dra: non-magnetic

CM Dra consists of two M4.5 dwarf stars in a 1.27 day eclipsing orbit. The system is viewed almost exactly (within less than 1 degree) in the orbital plane. The system has a space velocity of 165 km/sec (Lacy 1977), suggesting a Population II object. An age consistent with Population II can be derived from the presence of a common proper motion companion which is a white dwarf star (Morales et al. 2009): this leads to an age of 4.1± 0.8 Gyr.

### 1.2. Observational properties of CM Dra: starspots

Between eclipses, Lacy (1977) reported the presence of quasi-sinusoidal variations in brightness with amplitudes of ±0.021 mag on periods of 1.27 days. Since CM Dra is a close enough binary to ensure synchronization of each component with the orbital period (also 1.27 days), Lacy attributed the variations to the presence of starspots on the surface of one or both components.

Rotational modulation of the brightness of CM Dra has also been reported for the years 1996-2005 by Kozhevnikova et al (2009): the amplitudes were found to range from ±0.016 mag to ±0.026 mag. These results indicate that, over the course of some 30 years, some 1-3 percent of the surface area of CM Dra is periodically "blacked out" from the visible disk every rotational cycle. In view of the sinusoidal nature of these variations (i.e. the absence of flat-topped or flat-bottomed light curves), Lacy (1977) noted that the spots "must be located very close to the rotational poles". The significance of near-polar spots has recently been re-emphasized by Morales et al (2010). Homogeneous spots which are strictly confined above a certain latitude at all longitudes are not subject to rotational modulation in a system which is viewed equator-on (such as CM Dra). The 1-3% amplitudes in light must be due to lower-latitude "straggler spots" at certain longitudes. The percentage of the surface which is covered by homogeneous polar spots is not limited to 1-3%, but may be much larger.

By analogy with what we know about the umbrae of sunspots, it is plausible to attribute the occurrence of starspots in CM Dra to the presence of locally strong magnetic fields which are predominantly perpendicular to the surface (e.g. Mullan 1974): such fields, when they have horizontal scale-sizes in excess of one granulation cell, are effective at interfering with convective transport. (Horizontal fields have much less effect.)

**1.3. Observational properties of CM Dra: flares**

Also by analogy with the solar case, we expect that the presence of magnetic fields in a stellar atmosphere where convection is vigorous (such as in an M dwarf) should lead to occasional flare activity. Indeed, flares have been reported on CM Dra since 1966 (Eggen & Sandage 1967). Other reports of flares can be found in Lacy (1977), Vilhu et al. (1989), Kim et al. (1997), Deeg et al. (1998), Saar & Bookbinder (1998), Nelson and Caton (2007), and Kozhevnikova et al. (2009).

The flare energies in CM Dra are large compared to solar flares: in the R band alone, total energies range from $4 \times 10^{31}$ to $6 \times 10^{32}$ ergs (Kozhevnikova et al. 2009). Optical flare energies are typically strongest in the U band, and progressively weaker in the B and V bands (Gershberg 2002). Flare energies in the R band are expected to be even weaker than in the V band. The integrated flare luminosities in UBVRI are related to the luminosity in V by $L_{UBVRI} = 7.6 L_V$: the coefficient for $L_R$ should be even larger than 7.6. Thus, the total photon energies of flares on CM Dra may be some 10 times larger than the above R-band energies. Energies as large as even $10^{32}$ ergs are observed in only the very largest solar flares, and then only when all components of the energy (particles, photons, kinetic energies of ejecta) are included. Presumably, if we could evaluate these other components of flare energy in CM Dra, the total energies could exceed the largest solar values by $\geq$ (10-100). Now, flare energy scales as $VB^2$ where $V$ is the flare volume in which the mean field strength is $B$. A (10-100)-fold increase in $VB^2$ compared to the largest solar values would require stronger $B$ and/or larger $V$. With stellar surface areas in CM Dra A and B of only 5-6% of the solar value, an increase in $V$ may be difficult to achieve. An increase in $B$ by factors of $\geq$(3-10) compared to the Sun could suffice. In Section 8 below, we shall in fact find that our magnetic models point to poloidal field strengths on CM Dra which are $\geq$10 times solar.

Alternatively, an increase in $V$ could be handled if flares on CM Dra occupy a much larger fraction of the stellar surface than solar flares do. Is this a likely possibility? Perhaps: Saar & Bookbinder (1998) claim that active regions on CM Dra may occupy fractions of the visible disk area of 10's of percent, much larger than in the Sun. In this regard, we note that the emission lines of C IV 1550 Å and Mg II 2800 Å in CM Dra are close to the saturation limit for mid-M dwarfs (Vilhu et al. 1989), suggesting large areal coverage of the surface by active regions.

The flare data allow one to determine the rates at which flares occur in CM Dra: the rates range from a low of 0.011 per hour (in 1995: Deeg et al. 1998) to 0.057 per hour (Nelson & Caton 2007). These flaring rates are quite low compared to Population I flare stars with similar luminosity: there, the flare rates can be $\geq$2 per hour (Kim et al. 1997). Despite the rapid rotation of each star in CM Dra (equatorial rotational speeds of about 10 km/sec in both components), it appears that the components of CM Dra are less likely than Population I stars by factors of 30-200 (or more) to undergo flares. It is possible that the strong tides in CM Dra may reduce differential rotation: this could reduce the shear which helps to build up magnetic stresses prior to a flare.

Finally, X-rays from CM Dra indicate that the distribution of coronal material is not spherically symmetric (Gudel et al. 2005). In XMM data extending over more than one eclipse, both components show large inhomogeneities in longitude, indicating multiple distinct active regions. As a function of latitude, for both components, much of the plasma is seen at intermediate latitudes. Thus, the coronal plasma, as seen in X-rays, prefers to avoid the polar caps. This can be understood if in the large polar spots, the local fields are mainly vertical. In such cases, the plasma at high latitudes would be free to flow outward, perhaps guided to lower latitudes where it can be trapped and build up sufficient emission measure to become detectable.

Gudel et al. (2005) report coronal densities in the CM Dra components approaching $10^{10}$ cm$^{-3}$. Such high densities make it more difficult for the onset of fast magnetic reconnection (Mullan 2010): if flares are triggered by fast reconnection, this could also help to explain the low flare rates in CM Dra.

### 1.4. Magnetism in the two components of CM Dra: comparable

The time-scale required to synchronize the rotation of each component to the orbital period in a binary with period $P$ (days) is roughly $10^4 \, P^4$ in units of years (Zahn 1977). As a result, the components of CM Dra (with $P = 1.27$ days) are synchronized to the orbital period provided that the age of the system is at least $10^5$ years. In fact, based on properties of the white dwarf common proper motion companion, the age of CM Dra is estimated to be 4.1 Gyr, with an uncertainty of about 20% (Morales et al. 2009). This leads us to expect that both of the M dwarfs in CM Dra are tidally locked to the orbital period. Moreover, since their radii are equal (within 10%), the rotational velocities of both stars are essentially the same. As a result, we expect that dynamo activity should lead to comparable magnetic field strengths in both components. This suggests that in our magnetic modeling of CM Dra, we should not expect significant differences between the magnetic properties of the components A and B. (This is in contrast to younger binaries where, in the absence of synchronization, there may be marked differences between the magnetic properties of the two components: e.g. MacDonald & Mullan 2009 [MM09], Mullan & MacDonald 2010 [MM10]).

In summary, the observational evidence indicates that magnetic fields play a significant role in the physics of the CM Dra system. We shall find that we need to incorporate the effects of these fields in order to obtain fits (within the statistical errors) to the radii and luminosities of the CM Dra components.

### 2. Previous models

### 2.1. Non-magnetic models of CM Dra

Shortly after Lacy's (1977) pioneering work, CM Dra attracted attention as an interesting object for testing stellar structure calculations. Paczyński & Sienkiewicz (1984), citing the "considerable cosmological significance" of the helium abundance in CM Dra, demonstrated an adiabatic analysis which leads to values of $Y$ (the He mass fraction) for both components: they found $Y = 0.25 \pm 0.14$ and $0.38 \pm 0.12$ for CM Dra A and B respectively. Compared to the standard estimate of $Y = 0.270$ (Christensen-Dalsgaard et al. 1996) for the initial value for the Sun, these results hinted that perhaps CM Dra might have started its life with considerably high helium abundance than the Sun, perhaps as large as 0.5. However, the uncertainties were large. Follow-up analysis on more extensive data led Metcalfe et al. (1996), using the same technique, to improved uncertainties: they found $Y = 0.32 \pm 0.04$ and $0.31 \pm 0.04$ for A and B respectively.

Using a full stellar structure approach (rather than the adiabatic approximation), and relying on empirical radii which had uncertainties of 1 part in 30 (i.e. 4-5 times poorer than the current uncertainties), Chabrier & Baraffe (1995) obtained good fits to the data using $Y$ = 0.25-0.275 for both components. As a result, it seems unlikely that unusually large helium abundances are present in CM Dra. This will be relevant to our discussion in Section 6 below.

2.2. **Magnetic models: cool dwarfs in general**

In an early attempt to do "precision modeling" of M dwarfs with inclusion of magnetic fields, Cox et al. (1981) reported on their fit to the then-known parameters of the star Kruger 60A. This is a member of a binary which also includes the first flare star to have a known mass (Kruger 60B = DO Cep, with a mass of 0.14 $M_\odot$: Van de Kamp and Lippincott 1951). Kruger 60A has a mass of 0.26 $M_\odot$, not too different from those of CM Dra A or B. The biggest difference between Kruger 60 and CM Dra is the period: 44.52 years (Van de Kamp and Lippincott 1951) versus 1.27 days. Cox et al. (1981) found the mixing length ratio $\alpha = l/H_p$ in Kruger 60A had to be assigned values "in the range 0.07 to 0.17, rather than the more conventional values of 1.0 to 2.0". This is a dramatic reduction in $\alpha$ compared to standard models. Cox et al. (1981) suggested that "the low $l/H_p$ ratio is explained by the interaction of magnetic field with convection". The rotational velocity of Kruger 60A is small: $V \sin i \leq 3.0$ km/sec (Jenkins et al. 2009), so the magnetic field might not be too strong: the lack of any reported flaring activity in Kruger 60A (in contrast to Kruger 60B, with its faster rotation: $V \sin i$ = 4.7 km/sec, and detectable flare activity) is consistent with a relatively weak field in Kruger 60A. Despite this weak field, Cox et al. (1981) suggest that the field is enough to seriously impede convection, reducing $\alpha$ by an order of magnitude. Cox et al do not specify the quantitative relationship between the reduction in $\alpha$ and the magnetic field strength, but it must be a very sensitive function to cause a 10-fold reduction in $\alpha$ even in the presence of a rather weak field. One of our goals in this paper is to find a method (distinct from the reduced-$\alpha$ model) which allows us to make a quantitative estimate of the magnetic field strength on the surface of CM Dra, based on a particular physical model for the interaction between magnetic fields and convection.

As the presence of "bloating" and cooling of low-mass stars became better quantified, various quantitative explanations were presented in the literature. Mullan & MacDonald (2001; hereafter MM01) showed that, in the presence of magnetic fields which impede the onset of convection, models of low-mass stars undergo a simultaneous increase in radius and decrease in effective temperature compared to non-magnetic models of the same mass. The effects were shown to be consistent with the (less precise) parameters of active stars which were available at that time.

Other authors have also suggested that magnetic activity could be responsible for the observed bloating and cooling (Torres & Ribas 2002; Morales et al. 2009 and references therein). Chabrier et al (2007) suggest that larger stellar radii could result from a combination of two effects: (a) a reduction in the convective mixing length ratio significantly below typical values obtained by solar fitting and (b) coverage of a significant fraction of the surface area by cool spots.

## 2.3. Magnetic models: CM Dra

Morales et al. (2010) have argued that, since the masses of the CM Dra components are both smaller than that for the conventional transition to complete convection, effect (a) (in Section 2.2 above) leads to negligible changes in radius. As a result, they focus on applying effect (b). They conclude that unrealistically high spot coverages need to be assumed to reproduce the observations. They then consider how the presence of cool, dark spots influences the determination of radii from light curve analysis of eclipsing systems. To do this, they apply a generalized eclipse modeling code to synthetic light curves which include the effects of starspots. To produce the synthetic light curves, Morales et al assume that the starspots are present on both components, and exist over a wide range of stellar co-latitudes $\theta$. The spread of starspots on each stellar surface is assumed to follow a number of different distributions: (1) distributed with a probability which increases linearly in cos $\theta$ from the pole to 40°, has a plateau from 40° to 45°, and then the probability decreases linearly in cos $\theta$ between 45° and the equator; (2) distributed with a probability which is zero between the equator and $\theta = 70°$, increases linearly in cos $\theta$ from 70° to 25°, and then the probability decreases linearly in cos $\theta$ between 25° and the pole, and (3) uniformly distributed at all latitudes. Once the distribution is decided upon, individual spots are placed on the surface: each spot is assumed to be circular, with a radius of 10°, and a temperature equal to 0.85 times that of the unspotted photosphere. Random Gaussian noise of 1% of the flux is then added to the simulations to reproduce the typical scatter of light curves.

   Morales et al. then fit the synthetic light curves using the Wilson-Devinney code, including the effects of spots. They report "satisfactory fits with 1-3 spots on different components", although their Figure 5 suggests the presence of dozens of spots on each component.

   For the case of their spot distribution (2), they find that their fits systematically *over*estimate the sum of the radii of the components by 2%–6%. For distributions less concentrated to the pole, the deviations induced by spots are close to being random.

   Applying a 3% systematic *decrease* to the radii measured by Morales et al (2009) for CM Dra, Morales et al (2010) find that models with an effective spot coverage of 17% and a mixing length ratio $\alpha = 1$ match the observed radii.

**2.4. The approach in this paper**

In the present paper, we use our stellar evolution models in an attempt to fit the empirical radii of the two red dwarf components of CM Dra as reported by Morales et al (2009). First, in the context of "standard" (non-magnetic) models, we quantify how different compositions (including heavy elements and helium) affect the radii and luminosities of the two stars (Sections 4.1, 4.2). We also consider how the stellar properties in the "standard" models depend on choice of mixing length ratio (Section 4.3), and on putative errors in opacity (Section 4.4). In Section 5, we go beyond the standard models and consider models in which the presence of a magnetic field leads to changes in $L$ and $R$. In Section 6, we try various combinations of parameters, but in no case can we fit the radii of Morales et al (2009). In Section 7, we turn our attention to the systematically smaller radii reported by Morales et al (2010), after correction for the presence of polar spots. With these, we find that we *can* obtain acceptable fits. One way to obtain fits is to incorporate magnetic inhibition of convective onset: using results from that approach, we find that we can predict the strength of the (poloidal) fields on CM Dra.

**3. Methodology**

Our stellar evolution program has been described in MM10 and references therein. Here we give some details of the code relevant to the current study. Our code does not make use of separate envelope calculations. Instead a relaxation method is used to solve for the structure of star as a whole. One advantage of this approach is that a consistent value of mixing length ratio is used throughout the star. The equation of state is derived by minimization of a model Helmholtz free energy that, in addition to the translational free energy and internal free energy due to species with bound states (including the $H_2$ molecule), includes a hard sphere term to model pressure ionization, and Coulomb and quantum corrections term. Electrons are assumed to obey Fermi-Dirac statistics and all other particles are assumed to have Maxwellian velocity distributions.

      Since we are applying our models to precise observations of surface properties, we need to ensure that the surface boundary conditions are treated appropriately. We assign a small value of optical depth (typically $\tau_1 = 0.01$ or $0.1$) is assigned to the center of the outermost zone of the stellar model. The temperature at this point, $T_1$, is related to the effective temperature, $T_{eff}$, by the Eddington approximation, $T_1^4 = T_{eff}^4 (0.5 + 0.75\tau_1)$. The second surface boundary condition is $p\kappa = g\tau_1$, where $p$ is the pressure (including any contribution to the pressure from magnetic fields), $\kappa$ is the opacity, and $g$ is the surface gravity. Note that our use of the Eddington approximation is fundamentally different from that discussed in Chabrier & Baraffe (2000) and their earlier papers.

Chabrier & Baraffe implemented the Eddington approximation in a very particular way: they match the Eddington atmosphere to an interior solution at an optical depth of order 100. But this choice of matching point is inappropriate for stars with convective envelopes where much of the energy flux, even at the photosphere, is due to convection (for which the Eddington approximation is inapplicable). By contrast, in our models we use the Eddington approximation only to determine the temperature at the very small optical depth $\tau_1$ of our outer point of the star. At $\tau = 0.01$-$0.1$, the energy in CM Dra models is carried completely by radiation, so that the Eddington approximation is applicable.

To test our approach we have compared our models to corresponding Chabrier & Baraffe (2000) models with non-grey atmospheres. We find that our solar composition model of mass 0.2 solar masses, age 5 Gyr and mixing length ratio 1.0 has an effective temperature which is 20 K cooler than the Chabrier & Baraffe model. This relative temperature difference between our model and theirs is less than 0.6%. This is less than the typical uncertainties in observationally inferred $T_{eff}$ values of M dwarfs and is certainly less than those for CM Dra. From comparisons like this, we find that our models for M dwarfs and also brown dwarfs are essentially equivalent to those of Chabrier, Baraffe and coworkers in their surface properties. This gives us confidence that use of atmospheres in our models would not have a significant effect on the conclusions of this paper.

We calibrate the mixing length ratio, $\alpha$, by solar modeling. Our solar models have been constructed with and without including the effects of element diffusion (including thermal diffusion terms) and gravitational settling. We note here that we allow all the species to diffuse independently. However we find that this gives abundances that differ by less than 2% from the case in which all heavy elements are assumed to diffuse together. Based on the results of helioseismology (Christensen-Dalsgard et al 1996), we have assumed that for the present-day Sun the ratio of the heavy element mass fraction to the hydrogen mass fraction is $Z/X = 0.0245$. Our solar model with element diffusion and gravitational settling that matches the present day values of luminosity and radii at age 4.6 Gyr has $\alpha = 1.684$, initial helium mass fraction $Y_0 = 0.290$, and initial heavy element abundance $Z_0 = 0.0194$. We shall refer to these abundances as the protosolar abundances. The present day values of the mass fractions are found to be $Y_\odot = 0.261$, $Z_\odot = 0.0177$.

We then evolved models of stars of masses equal to those of the components of CM Dra assuming that the mixing length ratio is the same as our solar model and they have protosolar composition. Since the CM Dra models are fully convective, we did not include the effects of element diffusion and gravitational settling. In figure 1, we show the evolutionary tracks of the CM Dra models in the log $R/R_\odot$ - log $L/L_\odot$ plane. At early times in the evolution (<1 Gyr), the tracks move downwards and towards the left, reaching the main sequence at the lowest point on each track. At later times (up to 10 Gyr), each star evolves upwards and to the right along a track which can just barely be distinguished from the pre-main sequence track. The point on each track at the age of the proper motion companion, 4 Gyr, is marked by an ×. Also shown are points and error bars at the locations of the observed parameters found by Morales et al (2009). We have chosen to show our results in this plane, rather than the more usual Hertzsprung – Russell diagram (i.e. $L$ versus $T_{eff}$), because the high precision of the radius measurements means that the effective temperature is correlated with the luminosity.

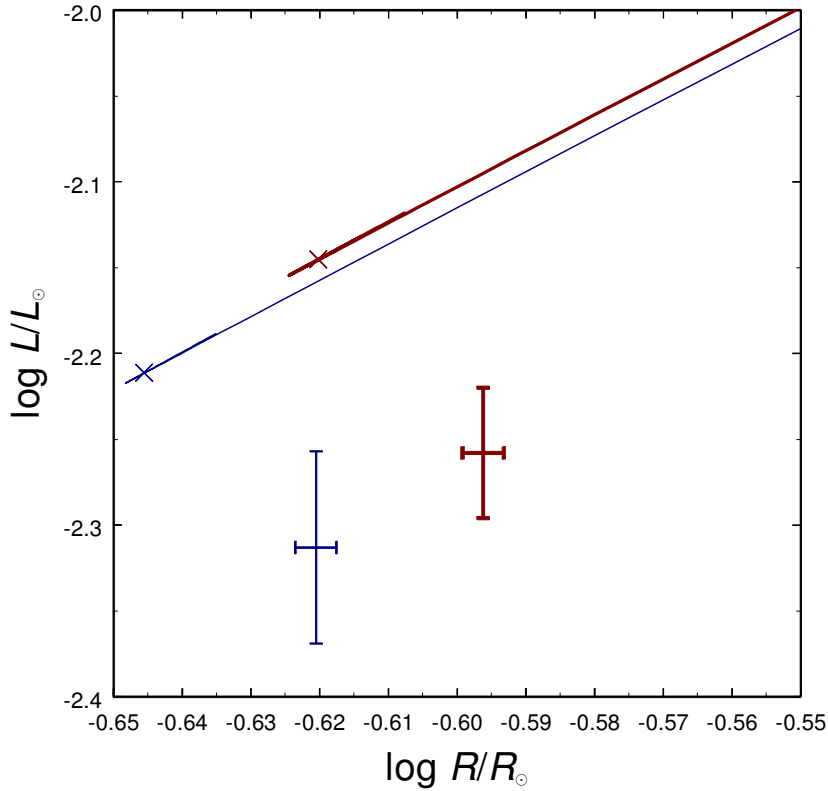

Figure 1. Standard (non-magnetic) evolutionary tracks for CM Dra A (red line) and CM Dra B (blue line) assuming solar mixing length ratio and primordial composition. The error bars are for the observational data of Morales et al (2009). The points on the tracks corresponding to age 4 Gyr are marked by ×.

We see that the evolutionary tracks do not pass through the statistical error boxes. At the radii of the stars, the models are too luminous by $4.2\sigma$ for A and $2.8\sigma$ for B. At the age of the proper motion companion, the models are too small by more than $7\sigma$.

We now consider various possible causes of the discrepancies between models and observations, beginning with consideration of how the values we adopt for various abundances lead to changes in the model properties.

## 4. Abundance effects

### 4.1. Heavy element abundances

To explore how the stellar properties depend on the adopted abundances we first consider models with different heavy element mass fractions, keeping $Y$ fixed. Figure 2 shows the evolutionary tracks for three values of $Z$: $Z_0/3$, $Z_0$ and $3Z_0$. We see that increasing $Z$ moves the tracks closer to the error boxes. However, the minimum value of $Z$ for which the tracks intersect the error boxes is $> 3Z_0$. It seems unlikely that in a Pop. II object such as CM Dra, the heavy element abundance could be so much larger than the solar value. The results in Fig. 2 suggest that we cannot fit the observed $L$ and $R$ of CM Dra by varying $Z$.

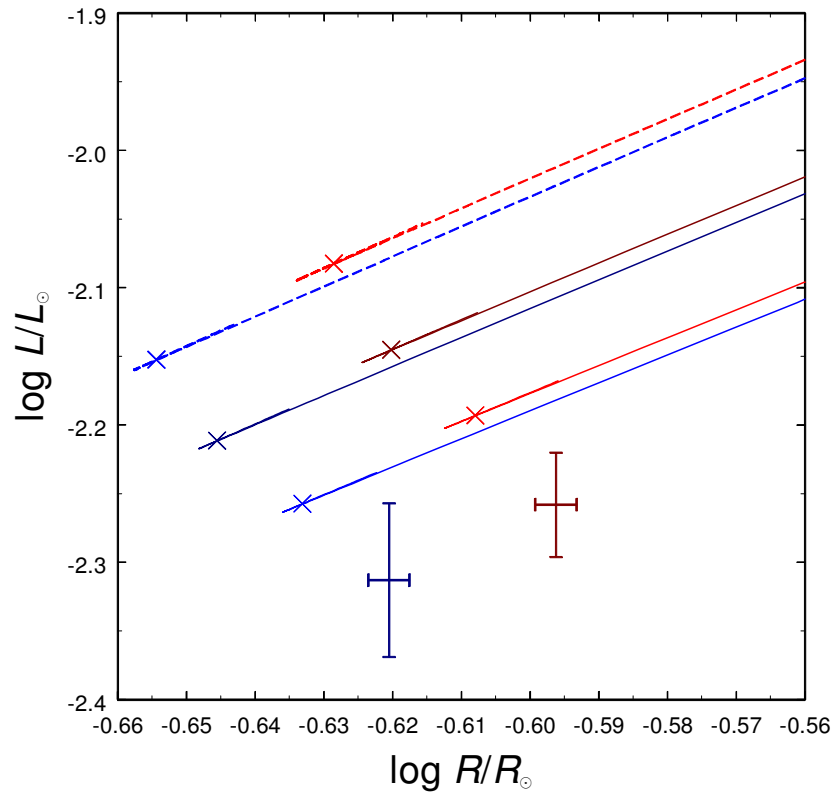

Figure 2. Non-magnetic evolutionary tracks for CM Dra A (red lines) and CM Dra B (blue lines) assuming solar mixing length ratio and primordial heavy element abundances $Z_0$ (dark lines), $Z_0/3$ (light, broken lines) and $3Z_0$ (light, solid lines). The error bars are for the observational data of Morales et al (2009). The points on the tracks corresponding to age 4 Gyr are marked by ×.

The effects of varying $Z$ are quantified by the logarithmic derivatives at $Z = Z_0$, $\partial \ln L/\partial \ln Z = -0.12$, $\partial \ln R/\partial \ln Z = 0.02$, and $\partial \ln R/\partial \ln L = -0.19$, which show that the luminosity is more strongly dependent on $Z$ than the radius.

**4.2. Helium**

We next consider the effects of varying the helium abundance, keeping $Z$ fixed. Figure 3 shows the evolutionary tracks for three values of $Y$: 0.2, $Y_0$ and 0.4. We see that decreasing the helium abundance moves the tracks to smaller luminosity and radius. But in order to fit the observed $L$ and $R$ values at any age, the $Y$ value would have to be smaller than 0.2, i.e. less than the cosmological limit of $Y_p = 0.249 \pm 0.009$ (Olive & Skillman 2004). Thus, changing the helium abundance by itself does not lead to permissible models which yield good fits to $L$ and $R$.

The derivatives at $Y = Y_0$ are $\partial \ln L/\partial Y = 0.96$, $\partial \ln R/\partial Y = 0.27$, and $\partial \ln R/\partial \ln L = 0.28$.

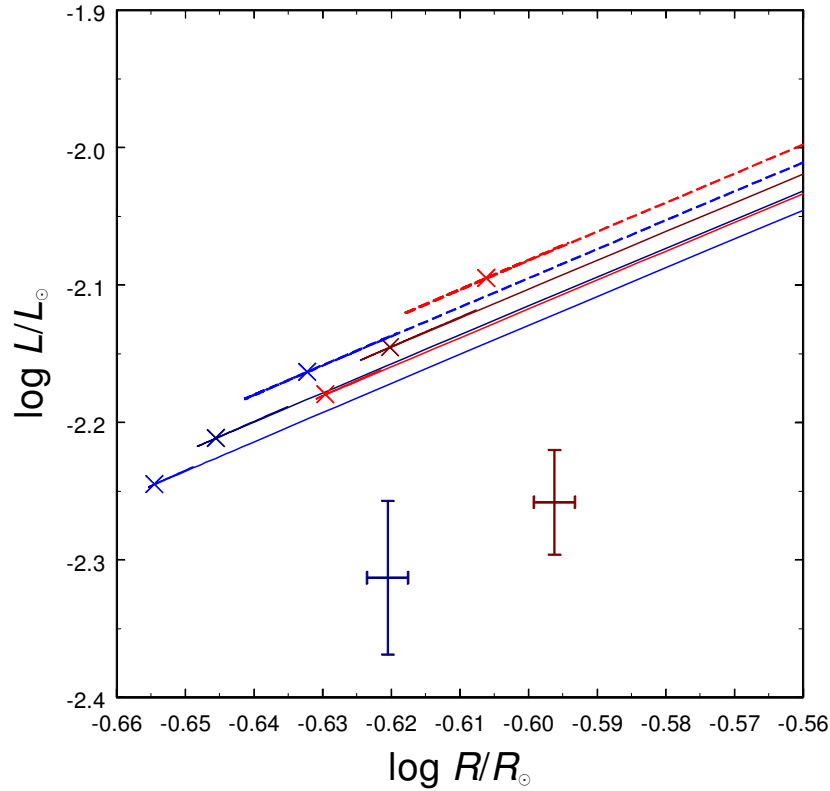

Figure 3. Non-magnetic evolutionary tracks for CM Dra A (red lines) and CM Dra B (blue lines) assuming solar mixing length ratio and helium abundances $Y_0$ (dark lines), 0.4 (light, broken lines) and 0.2 (light, solid lines). The error bars are for the observational data of Morales et al (2009). The points on the tracks corresponding to age 4 Gyr are marked by ×.

### 4.3. Mixing length ratio

In view of the results of Cox et al. (1981), we explore in Figure 4 how the evolutionary tracks for the CM Dra components behave when we vary the mixing length ratios, $\alpha = 0.4, 0.8, 1.2$ and $1.684$. We see that reducing the mixing length ratio moves the tracks to larger radii and lower luminosity. The tracks with $\alpha = 0.4$ intersect the error boxes for CM Dra A and B. However the intersections do not occur *on the main sequence*, but rather on the *pre-main sequence* phase of evolution. It is difficult to see how the age of the CM Dra system (4 Gyr) can be reconciled with pre-main sequence status. At age 4 Gyr, the $\alpha = 0.4$ models are discrepant in radius (too small by 3-4%) to agree with the observations, although the $L$ values agree well with the observations. We shall return below (Section 7) to the issue of radius discrepancies of about 3%.

At age 4 Gyr, the derivatives of radius and luminosity are $\partial \ln L / \partial \alpha = 0.08$, $\partial \ln R / \partial \alpha = -0.007$, and $\partial \ln R / \partial \ln L = -0.09$, which shows that a change in mixing length has an order of magnitude larger effect on luminosity than on radius. This insensitivity of $R$ to changes in $\alpha$ was remarked on by Morales et al. (2010).

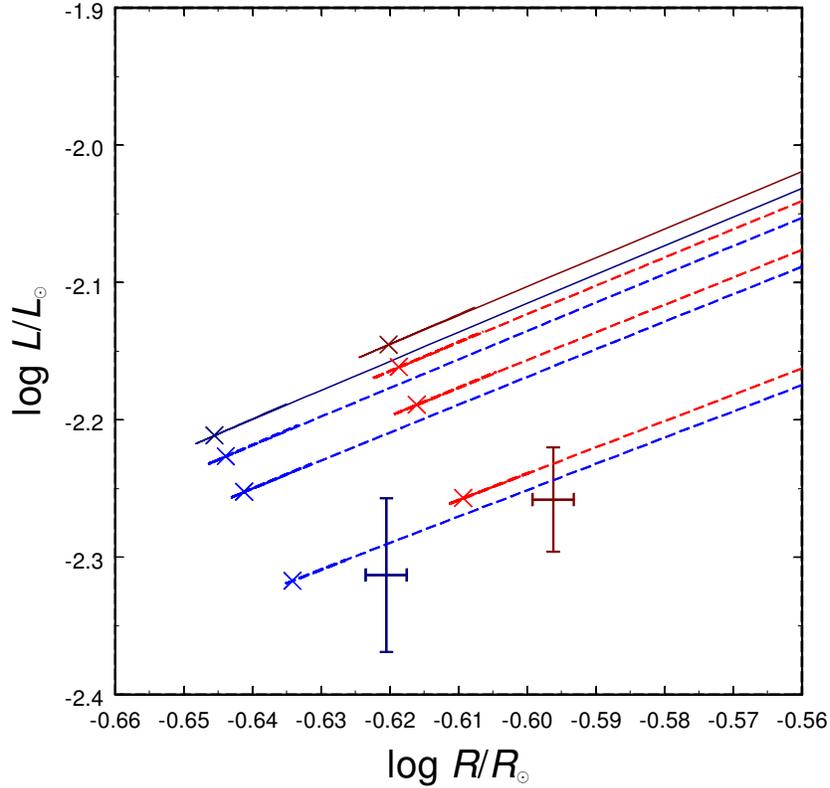

Figure 4. Non-magnetic evolutionary tracks for CM Dra A (red lines) and CM Dra B (blue lines) assuming primordial solar composition, and 4 values of $\alpha$. The tracks for a solar value of $\alpha$ are shown by dark lines. The light broken lines in order from the top are for $\alpha = 1.2$, 0.8 and 0.4. The error bars are for the observational data of Morales et al (2009). The points on the tracks corresponding to age 4 Gyr are marked by ×.

## 4.4. Opacity

In a sample of M1.0-M3.0V stars where angular radii have been measured by the CHARA array, Berger et al. (2006) found that the measured stellar radii exceed the theoretical models by up to 15-20%. Berger et al. noted that heavy element abundance may play a significant role in the discrepancies: the differences were observed to be smallest among metal-poor stars. Berger et al. suggest "that theoretical models for low-mass stars may be missing some opacity source that alters the computed radii".

To study the effects of increasing the opacity we have calculated a model of CM Dra A in which the radiative opacity is uniformly increased by an arbitrary factor $f_\kappa = 10$. The resulting track is shown by the thick green line in Fig. 5, where we also plot results for various values of $\alpha$ (1.684, 0.4, and 0.2 from top to bottom). In the main sequence phase, the models indicate that increasing the opacity is essentially equivalent to reducing the mixing length ratio. Once again, the high-opacity models with age 4 Gyr are not consistent with the observed $L$ and $R$ within the error bars. An independent argument which suggests that missing opacity is not likely to be the cause of oversizing in fully convective lower main sequence stars is provided by Lopez-Morales (2007): she reports that although the fractional discrepancies in radii among low-mass stars reach values as large as 30% among M dwarfs with masses in the range 0.4 - 0.6 $M_\odot$, in the fully convective stars with masses of 0.3 $M_\odot$ and below, the fractional discrepancies in radii become smaller.

The derivatives at age 4 Gyr are $\partial \ln L/\partial \ln f_\kappa = -0.4$, $\partial \ln R/\partial \ln f_\kappa = 0.04$, and $\partial \ln R/\partial \ln L = -0.1$.

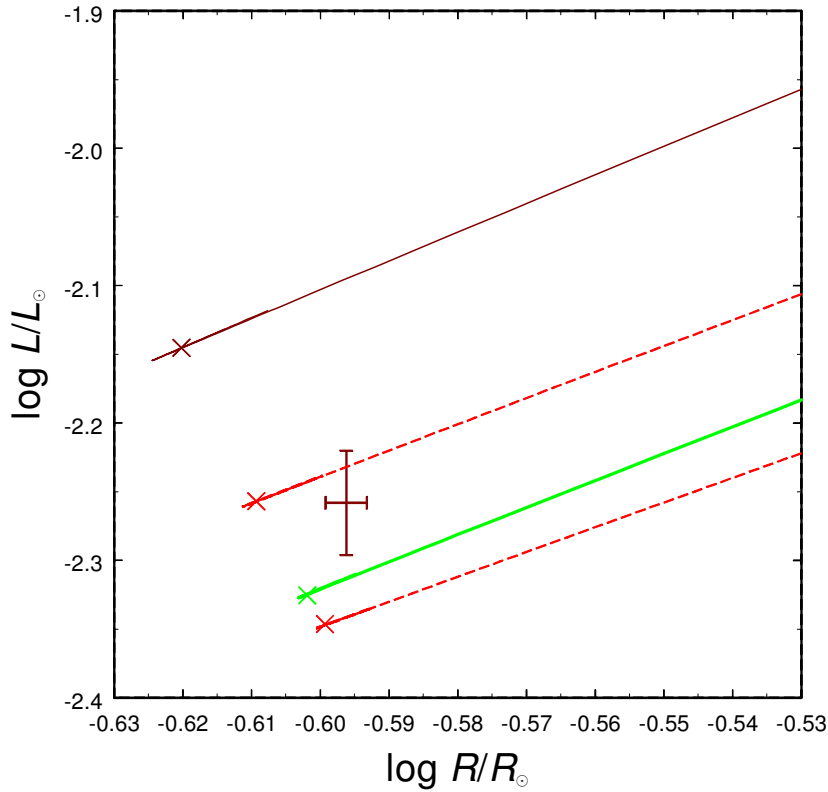

Figure 5. Thick green line: non-magnetic evolutionary track for a model of CM Dra A with primordial solar composition and solar value of $\alpha$ but with the radiative opacity increased (arbitrarily) by a factor 10 throughout the star. Also shown by red lines are evolutionary tracks assuming primordial solar composition for 3 values of $\alpha$. The track for a solar value of $\alpha$ is shown by the dark red line. The two broken light red lines in order from the top are for $\alpha = 0.4$ and 0.2. The error bars are for the observational data of Morales et al (2009). The points on the tracks corresponding to age 4 Gyr are marked by ×.

## 5. Models which include effects of magnetic fields on convection

Cox et al. (1981) interpreted their results of fits to Kruger 60A (in which they concluded that the mixing-length parameter α must be very small [0.07-0.17]: see Section 2.2) in terms of some kind of magnetic interaction with convection. However, Cox et al. offered no specifics about this interaction. In particular, they offered no physically-based formula which relates the reduced value of α to the local magnetic field strength. In order to be more quantitative about the effects which magnetic fields have on convection, MM01 incorporated two effects which are not generally included in the standard code of (non-magnetic) stellar evolution. First, the radial gradient of magnetic pressure was included in the equation of hydrostatic equilibrium. Second, a particular criterion was applied to allow for the physical fact that, in the presence of a magnetic field, the *onset* of convection is inhibited by a quantifiable amount compared to the non-magnetic case. The quantitative nature of the inhibition gives us an important possibility: deriving a value for the local magnetic field strength.

### 5.1. Criteria for the onset of convection in the presence of a magnetic field

The criterion which was used by MM01 for onset of magneto-convection was derived by Gough & Tayler (1966: GT). GT showed that, in the presence of a vertical magnetic field $B_v$ in a perfectly conducting medium, the Schwarzschild criterion for the onset of (non-magnetic) convection, namely, $\nabla > \nabla_{ad}$ (where $\nabla = \partial \log T / \partial \log P$, $P$ is the gas pressure, and subscript *ad* refers to adiabatic conditions) is replaced with the following expression:

$$\nabla > \nabla_{ad} + \delta \quad (1)$$

where

$$\delta = \frac{B_v^2}{B_v^2 + 4\pi\gamma P} \quad (2)$$

In the limit of weak magnetic fields, such as those we discuss in the present paper, the quantity $\delta$ can be written with acceptable precision in the form

$$\delta \approx B_v^2 / 4\pi\gamma P . \quad (3)$$

We refer to $\delta$ as the "magnetic inhibition parameter".

It is important to note that the field component which inhibits convection, and which appears in eqs. (2) and (3), is the *vertical* component. Horizontal fields are less likely to interfere with the onset of convection, although the preferred planform may undergo significant alteration (from hexagons to rolls) (Chandrasekhar 1961).

In our first study of magnetic inhibition in stars (MM01), we applied the GT criterion to steady magnetic fields in stars on the lower main sequence. We found consistency with the ("bloated") radii and (reduced) effective temperatures which were available at that time for stars with the highest levels of magnetic activity. In more recent work, the effects of constant fields on stellar evolution has been extended to brown dwarfs (MM09, MM10). In the latter works, the GT criterion was generalized to include the effects of finite electrical conductivity: in the outer layers of very cool dwarfs, where appreciable concentrations of neutral gas may exist, magnetic fields can diffuse rapidly through the medium. As a result, the field no longer interferes as effectively with convection. Using the results of Chandrasekhar (1961), the GT criterion was generalized to a criterion which we labeled GT-C: this includes the case of finite magnetic diffusivity.

**5.2. The choice of radial profile of $\delta$**

The most significant unknown in attempts to apply the GT (or GT-C) criteria to actual stars has to do with choosing the radial profile of $\delta$ between the center of the star and its surface. It may eventually become possible, using the observed changes in p-mode frequencies in the Sun between solar minimum and solar maximum (e.g., Mullan et al. 2007 [MMT07]) to derive some information on the radial profile of $\delta$ in the solar convection zone. But such information is not likely to become available for CM Dra for a long time (if ever).

In view of our ignorance of $\delta(r)$, the best we can do is to consider some general ideas. Various functional forms for the radial profile of $\delta$ can be envisioned.

(i) The simplest choice of radial profile is $\delta(r) = \delta$ = constant throughout each star (e.g. MM01; MM09, MM10).

(ii) Allow $\delta$ to decrease smoothly as we go into the star, with the largest value at the surface, and $\delta \rightarrow 0$ at the center of the star: MM01 found that, by using the functional form $\delta(r) \sim m(r)^{2/3}$, the results did not differ greatly from those with $\delta$ = const.

(iii) In the Sun, where time-variable fields were being modeled, MMT07 set $\delta$ = const in the convection zone ($r > 0.7R(sun)$). However, for physical reasons, $\delta(r)$ is required to decrease exponentially to zero over a narrow range of distances at the top of the radiative interior. This behavior was determined by the very high electrical conductivity of the radiative interior, which forbids short-term variability in field strength. This option is not relevant in the present paper, where we deal with a convective star (where turbulence leads to low effective electrical conductivity), and time-invariant fields.

(iv) In the present paper, for reasons which will be discussed in Section 8 below in the context of dynamo models of convective stars, we examine a "ceiling option": we keep $\delta$ = const from the surface down into the star until we reach a radial location $r_c$ at which the local field strength takes on a certain value (the "ceiling" field strength). In the deeper layers with $r < r_c$, we fix *the magnetic field strength* constant at its ceiling value. As a result, in the deep interior, $r < r_c$, $\delta(r)$ decreases to small values, although it never falls to zero. The physical basis for a "ceiling model" has to do with the ability of a dynamo to generate fields: a dynamo can generate fields only up to a certain limiting strength. The limit occurs when the magnetic energy density rises to values which are (roughly) in equipartition with the kinetic energy density (Browning 2008). We shall argue below (Section 8) that the maximum field strength in CM Dra may be of order $10^6$ G.

In our previous studies of certain stars (MM09, MM10), we considered considerably stronger fields than this (10-100 MG): the difference between those previous studies and the present one is that MM09 and MM10 were dealing with stars which are much younger than CM Dra. In MM09, the age of the system, was only 1-2 Myr, while in MM10, the age was of order 0.3 Gyr. In such stars, primordial fields may still be present (to some extent), in which case limits associated with dynamo action are less relevant. But in CM Dra, with an age of 4-5 Gyr, it is much more likely that fields in the star are associated with a dynamo. As a result, we will eventually (Section 7.3.3 below) consider models where the field strength is not allowed to exceed 1 MG.

### 5.3 Results for CM Dra

We first consider models in which magneto-convection is described by the GT criterion, and we use radial profile option (i), i.e. $\delta$ = const throughout the star.

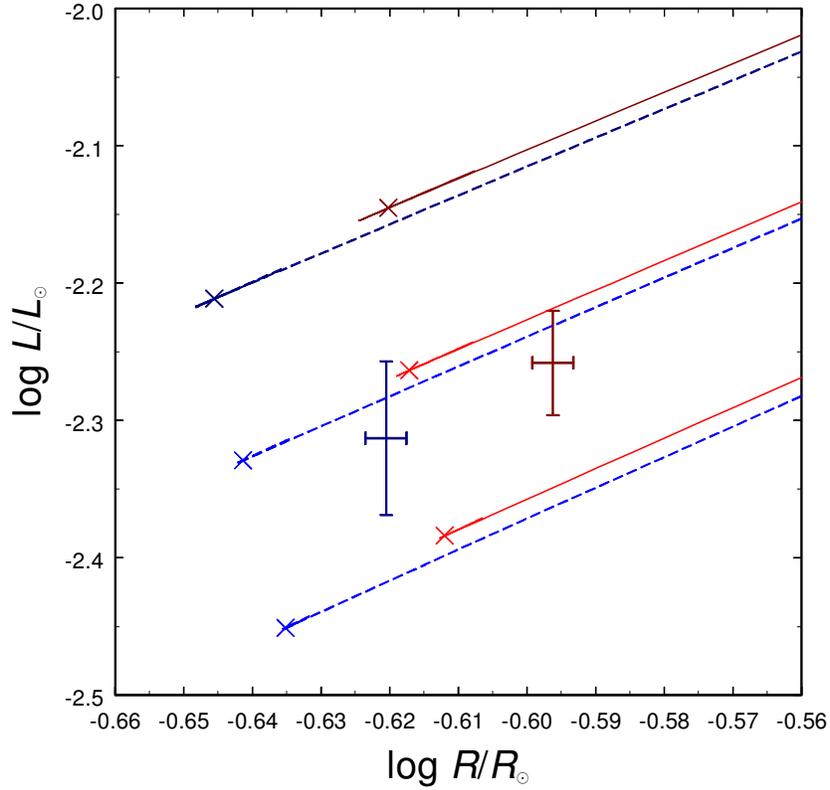

Figure 6. Evolutionary tracks of magnetic models of CM Dra A (in red) and CM Dra B (in blue) using the GT criterion for magnetic inhibition for 3 values of the (radially constant) magnetic inhibition parameter: $\delta = 0$ (dark lines), 0.01 and 0.02. Also shown are the error bars for the observed radii and luminosities from Morales et al (2009). The points on the tracks corresponding to age 4 Gyr are marked by $\times$.

Figure 6 shows the evolutionary tracks for 3 values of the magnetic inhibition parameter: $\delta = 0$, 0.01 and 0.02. Magnetic inhibition causes the tracks to move to lower luminosity, with only a small increase in radius. The tracks with $\delta = 0.01$ pass through the error boxes, but as with the reduced $\alpha$ models, during the pre-main sequence phase of evolution. The derivatives are $\partial \ln L/\partial \delta = -27$, $\partial \ln R/\partial \delta = 0.7$, and $\partial \ln R/\partial \ln L = -0.03$, which shows that magnetic inhibition has a much larger effect on luminosity than radius, as was the case for changes in mixing length ratio.

We next consider models calculated with the GTC model for magnetic inhibition of convection, again using radial profile option (i). Figure 7 shows the evolutionary tracks for 5 values of the magnetic inhibition parameter: $\delta = 0$, 0.02, 0.04, 0.06 and 0.08. As for the GT model, magnetic inhibition moves the tracks to lower luminosity but in contrast the radii decrease by a small amount. This subtle effect is due to magnetic inhibition causing the inner parts of the star to expand slightly. The outer layers are then slightly cooler and the degree of ionization is less. This reduces the inhibition of convection there.

The derivatives are $\partial \ln L/\partial \delta = -9.7$, $\partial \ln R/\partial \delta = -1.1$, and $\partial \ln R/\partial \ln L = 0.1$, which shows that a change in $\delta$ again has an order of magnitude larger effect on luminosity than on radius.

Whether we consider the GT or the GT-C criterion for magnetic inhibition, our models do *not* fit the observed $L$ and $R$ within the statistical error bars for an age of 4 Gyr.

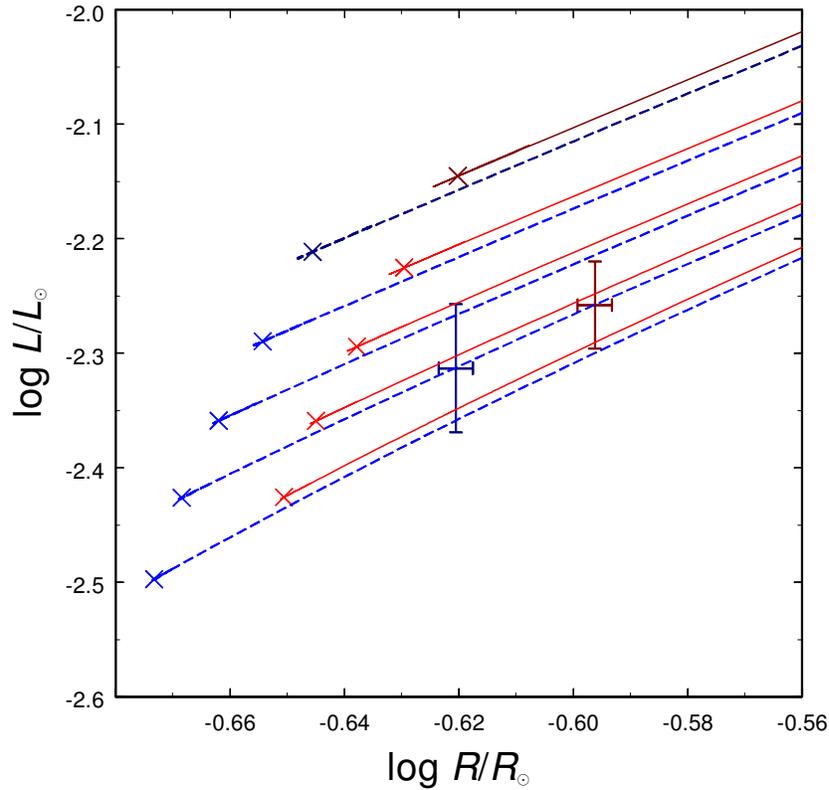

Figure 7. Evolutionary tracks of magnetic models of CM Dra A (in red) and CM Dra B (in blue) using the GTC criterion for magnetic inhibition for 5 values of the (radially constant) magnetic inhibition parameter: $\delta = 0$ (dark lines), 0.02, 0.04, 0.06 and 0.08. Also shown are the error bars for the observed radii and luminosities from Morales et al (2009). The points on the tracks corresponding to age 4 Gyr are marked by ×.

## 6. Combined effects

Although none of the individual effects considered above permit fitting of the observed $L$ and $R$ values of the CM Dra components by the models at an age of 4 Gyr (equal to that of the proper motion companion), we now explore whether a combination of helium abundances, heavy element abundances, and reduction of efficiency of convective energy transport, either by choosing a low value of $\alpha$ or by magnetic inhibition of convection, can. To determine the region of $Y - Z$ parameter space for which CM Dra A can be fitted at the assumed age, we have calculated additional evolutionary tracks for combinations of $Y$, $Z$ and $\alpha$ that span the ranges $0.2 < Y < 0.4$, $-0.5 < \log Z/Z_0 < 0.5$, and $0.2 < \alpha < 1.684$. Figure 8 shows the region of $Y$, $Z$ space for which fits can be found by adjusting $\alpha$. The colors show the maximum value of $\alpha$ for which a fit is possible. We see that for protosolar heavy element abundance, a fit is possible only if $Y > 0.34$. For $Y = 0.34$, the mixing length ratio is $\alpha = 0.256$. Fits with larger $\alpha$ can be found by increasing $Y$. For example, at $Y = 0.4$ fits occur provided $0.270 < \alpha < 0.382$.

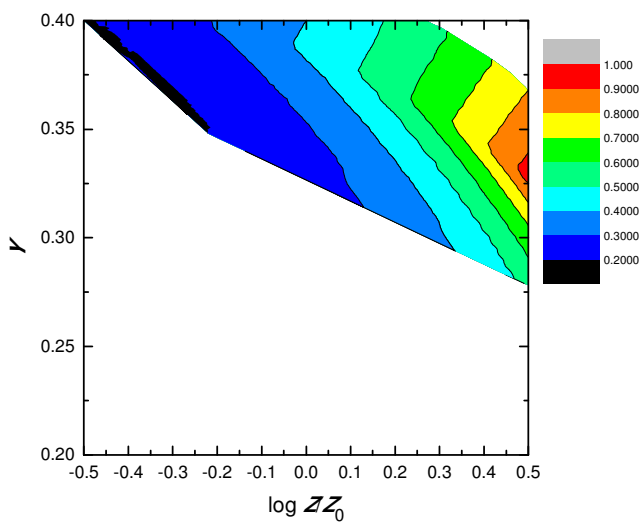

Figure 8. The region of $Y - Z$ space for which fits of non-magnetic models to the observed properties of CM Dra A can be found by adjusting $\alpha$. The color table is labeled with values of $\alpha$.

Figure 9 is the same as for figure 8, except that we now show the region of $Y$, $Z$ space for which fits can be found by adjusting $\delta$ in the GT models. (In Fig. 9, we refer only to models with radial profile option (i), i.e. $\delta(r) =$ const.) Note that the axis ranges are shorter than in figure 8. Fits with protosolar heavy element abundances require $Y$ greater than the largest value that we considered, $Y = 0.40$. Similarly fits with protosolar helium abundance require $Z$ greater than 3 times the protosolar value.

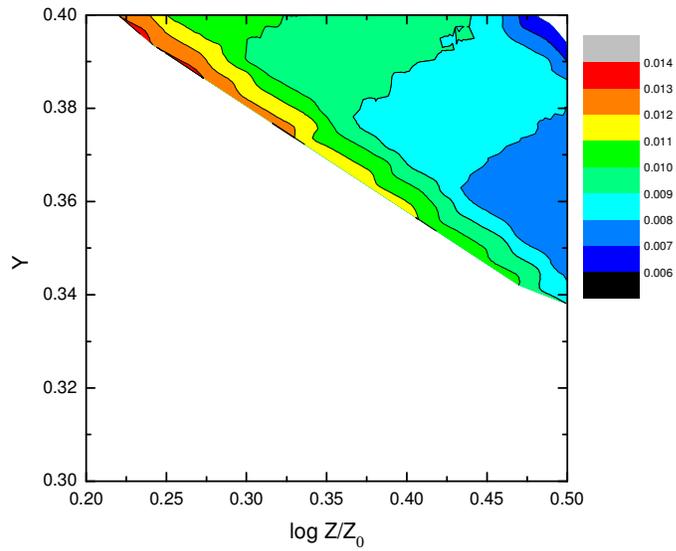

Figure 9. The region of *Y* - *Z* space for which fits of magnetic models to the observed properties of CM Dra A can be found by adjusting $\delta$. The GT criterion is used for the onset of convection in the presence of a magnetic field. The color table is labeled with values of $\delta$. All cases in this figure have $\delta$ = const with radius.

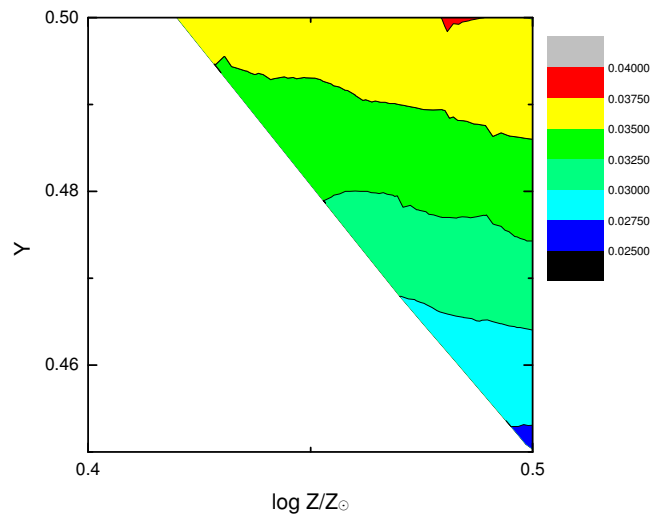

Figure 10. The region of *Y* - *Z* space for which fits of magnetic models to the observed properties of CM Dra A can be found by adjusting $\delta$. The GTC criterion

is used for the onset of convection in the presence of a magnetic field. The color table is labeled with the values of $\delta$. All cases in this figure have $\delta$ = const with radius.

Figure 10 differs from figure 9, in that we now show the region of $Y$, $Z$ space for which fits can be found by adjusting $\delta$ in the GTC models. To find fits with $Z$ in the range considered, we need to consider additional models calculated with $Y = 0.5$. Fits with protosolar heavy element abundances were not possible for $Y < 0.50$. Also fits with protosolar helium abundance were not found for $Z$ less than 3 times the protosolar value.

We conclude that for heavy element abundances in the range 0.1 to 1 times solar, fitting models of the CM Dra components at age 4 Gyr to the global properties found by Morales et al (2009) requires a significantly larger than solar helium abundance. The models with the lowest helium abundance, $Y = 0.34$ require solar heavy element abundance and a mixing length ratio, $\alpha = 0.256$, much lower than the value obtained from solar fitting, $\alpha = 1.684$. And even the lowest of these $Y$ values is too large to be consistent with the values (≤0.275) reported by Chabrier & Baraffe (1995).

In the unlikely event that the white dwarf which shares a common proper motion with CM Dra (Morales et al. 2009) is a chance coincidence, there might be no age constraint on CM Dra at all. In such a case, an alternative scenario is that the components of CM Dra are in the pre-main sequence phase. A reduction of convective efficiency is still required but the data *can* be fit in this case without the need for a super-solar helium abundance. For example, a primordial composition model with $\alpha = 0.4$ fits CM Dra A at age 0.24 Gyr. Similarly fits are obtained for GT and GTC primordial composition magneto-convection models with (radially constant) $\delta = 0.0125$ at age 0.21 Gyr, and $\delta = 0.06$ at age 0.18 Gyr, respectively.

**7. Models which include the effects of spots**

Morales et al (2010) have studied how the presence of spots on the surface of a star can affect two of the properties (surface temperature, radius) which one derives for the star using standard eclipse modeling.  They report the important result that the presence of polar spots (such as Lacy [1977] mentioned in connection with CM Dra) leads to a systematic *over*-estimation of the radii by the eclipse modeling program. They find that this effect could give rise to a systematic error in radius determination of up to 6%. Here we summarize separately the effects which spots have on determinations of $T_s$ and $R$.

**7.1. Effects of spots on surface temperatures**

The influence of starspots on internal stellar structure has been reviewed by Spruit (1992). The blocking effects of spots is modeled by modifying the surface boundary condition to

$$L = 4\pi R^2 (1-f) \sigma T_s^4, \qquad (4)$$

where $f$ is the effective fraction of the surface covered by spots (assumed to be completely dark), and $T_s$ is the surface temperature of the immaculate (unspotted) surface. For fully convective stars, spots result in a reduction in luminosity given by $\Delta L/L \approx -f$, with much smaller relative changes in $R$ and $T_s$. Note that because $T_s$ does not change significantly, any significant spot coverage will reduce the effective temperature of the star. Hence to avoid confusion it is important to carefully distinguish between $T_s$ and $T_{eff}$ for heavily spotted stars.

Thermally relaxed models of spotted main sequence stars have been made by Spruit & Weiss (1986). They find that the effects of spots on energy transport in the convective zone are equivalent to an increase in Rosseland mean opacity in the outer part of the stellar envelope. Since fully convective lower main sequence stars are isentropic except for the outer layers, increasing opacity in the interior has little effect on the global structure. Hence the effects of spots on fully convective stars are equivalent to increasing the opacity uniformly throughout the star. Now, we have already noted (Section 4.4) that increasing the opacity in fully convective lower main sequence stars has the same effect as reducing the missing length. As a result, we expect that adding spots will be equivalent to reducing the mixing length. We have verified this by including spots in our models using equation (4). The resulting derivatives for our CM Dra models are $\partial \ln L/\partial f = -0.75$, $\partial \ln R/\partial f = 0.07$, and $\partial \ln R/\partial \ln L = -0.09$, which are consistent with the findings of Spruit & Weiss (1986) and Spruit (1992). The numerical value of $\partial \ln R/\partial \ln L$ is found to be the same as for reducing the mixing length. Hence for the theoretical models of fully convective stars, there is a degeneracy between the effects of reducing the mixing length ratio and including dark spots according to equation (4).

**7.2. Effects of spots on empirical estimates of stellar radii**

For the specific case of CM Dra, Morales et al (2010) find that, for their choice of spot distribution (2), i.e. spots concentrated towards the polar regions, an effective spot coverage of $f = 0.17$ reduces the observationally determined radii on average by 3%. This is a troubling result: it indicates that systematic effects may be several times larger than the statistical errors in the radii reported by Morales et al. (2009). Therefore, when we were trying to fit the $R$ and $L$ values of Morales et al. (2009) in Sections 4, 5, and 6 above, we may actually have been "aiming at the wrong target". The fact that we were unable to identify an acceptable fit now seems more understandable. Moreover, in a quantitative sense, we recall that some of our models turned out to have radii which were smaller than the "observations" of Morales et al (2009) by 3-4% (Section 4.4): this is precisely the amplitude of the reductions in radius which we must now incorporate into the fitting attempts.

**7.3. Modeling CM Dra with spot-reduced radii**

Once it is recognized that the previous estimates of the radius of CM Dra A may be systematically too large by several percent due to the effects of polar spots, we now revisit the "precision modeling" efforts which were reported in Sections 3-6 above. In this section, we attempt to fit the new (smaller) estimate of stellar radius which includes the correction for polar spots.

7.3.1. Mixing length variations

With empirical radii $R$ reduced by some 3%, Morales et al (2010) find that models which include spot coverage $f = 0.17$ along with $\alpha = 1.0$ can lead to agreement between theory and observations as far as radius is concerned. However from their figure 8 it seems that they have not adjusted $T_{eff}$ for the change in $R$: instead they seem to have fitted to the photometrically determined (non-spotted) $T_{eff}$ which was given by Morales et al (2009). This inconsistency means that the luminosity of their fitted models is below the observationally determined luminosity by ~7%.

Using the reduced $R$ value from Morales et al (2010), we find that an unspotted model for CM Dra A gives a good fit at age 4 Gyr for protosolar composition and $\alpha = 0.4$ (see Section 4.3 and Figure 4 above). For spot coverage $f = 0.17$, the same result is obtained for $\alpha = 0.63$. For the solar value, $\alpha = 1.684$, a precise fit is obtained for $f = 0.32$. Hence the Morales et al (2010) approach to modeling CM Dra can be successful by using a combination of spots and a significant reduction of convective efficiency from the solar value. (However, we do not need to reduce $\alpha$ as dramatically as Cox et al. (1981) found it necessary to do for Kruger 60A.)

Figure 11 shows the region of the $\alpha - f$ plane for which our CM Dra A models at age 4 Gyr are in agreement with the polar-spot-reduced radii within the statistical error boxes. Here we have assumed that the systematic shift in the observed radii due to polar spots is proportional to the effective fractional spot coverage. For $f = 0.17$ (the coverage preferred by Morales et al 2010), we can find agreement for $\alpha$ in the range 0.44 to 1.01. Note that our models can actually obtain fits *without reducing α below the solar value at all* if the spot coverage $f$ is as large as 0.225-0.245.

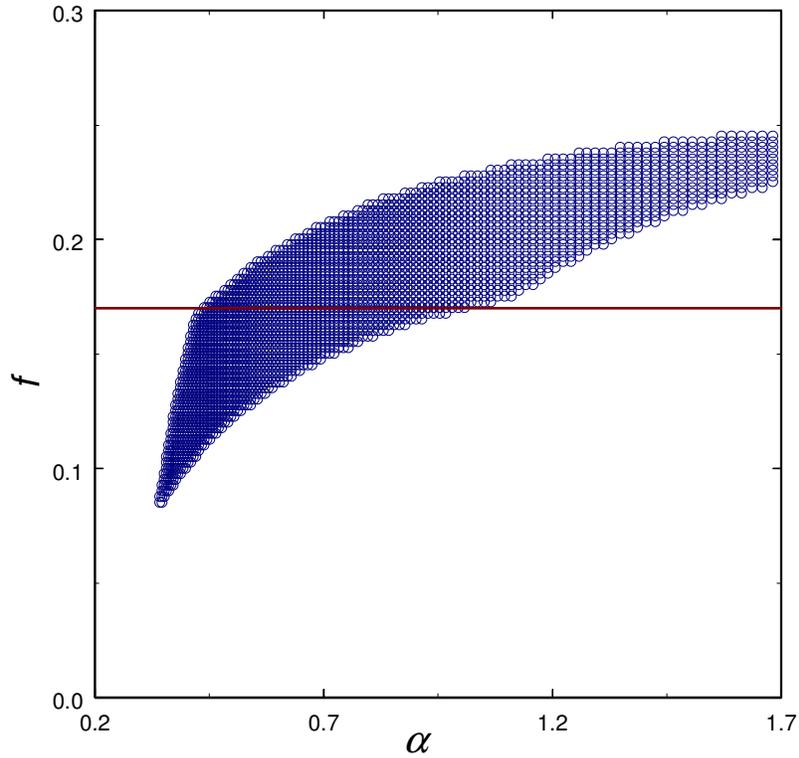

Figure 11. The region of the $\alpha - f$ plane for which non-magnetic models of CM Dra A have evolutionary tracks which pass through the error boxes. The systematic shift in the observed radii due to spots is assumed proportional to the effective fractional spot coverage, $f$. The horizontal line is placed at $f = 0.17$ (Morales et al. 2010).

7.3.2. Modeling the onset of magneto-convection in CM Dra: constant $\delta$

We have applied our magneto-convection models to modeling CM Dra A. We find that we can obtain fits to the reduced $R$ value from Morales et al. (2010) for certain combinations of the two parameters $f$ (see eq. (4)) and $\delta$ (see eq. (3)). In this sub-section, we report on results from models in which we use radial profile option (i) (see Section 5.2), i.e. $\delta$ = const with radius. The region of the $\delta$ - $f$ plane for which fits to the reduced $R$ value can be obtained with this option is shown in figure 12.

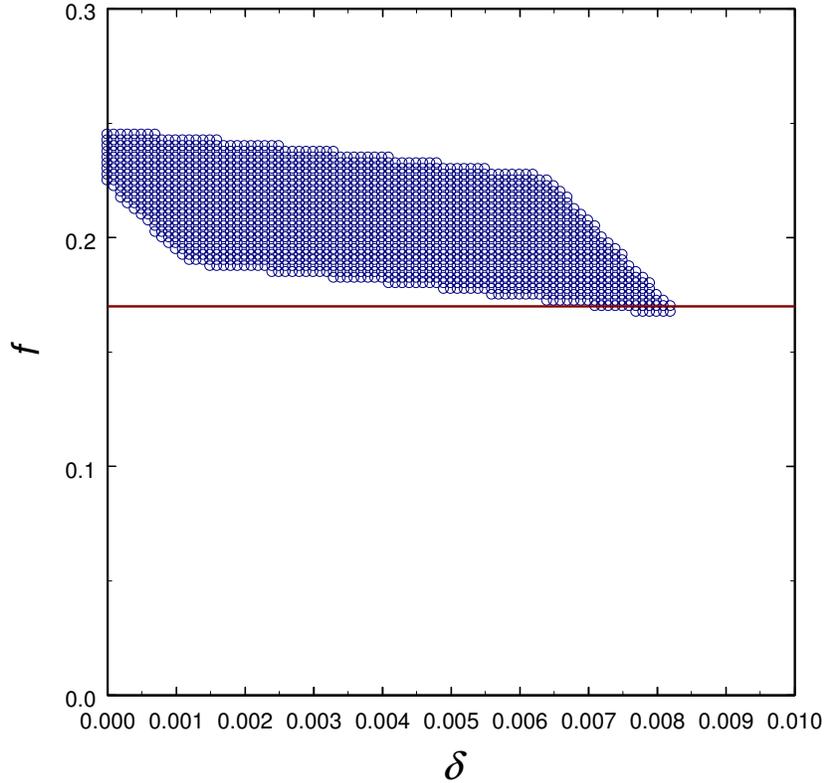

Figure 12. The region of the $\delta - f$ plane for which magnetic models of CM Dra A have evolutionary tracks which pass through the error boxes. The GT criterion has been used for the onset of convection in the presence of magnetic field, with $\delta =$ const at all radial locations. The systematic shift in the observed radii due to spots is assumed proportional to the effective fractional spot coverage, $f$. The horizontal line is placed at $f = 0.17$ (Morales et al. 2010).

Assuming a spot coverage $f = 0.17$, GT models with (radially constant) $\delta$ in the range 0.007 to 0.008 agree with the observations at age 4 Gyr. In view of the constraint $\delta =$ const, our solutions include fields which become very strong near the center of the star, of order $10^7$ G. We will return to a discussion of this result in Section 8.

Figure 13 is the same as figure 12, except that the GTC criterion is used for magnetic inhibition, and again $\delta =$ const at all radial locations. The minimum spot coverage for which a fit is found is $f = 0.20$. The largest values of $\delta$ which provide acceptable fits at age 4 Gyr are $\delta = 0.015$-$0.016$. These correspond to fields of order $10^7$ G at the center of the star. We will return to a discussion of this result in Section 8.

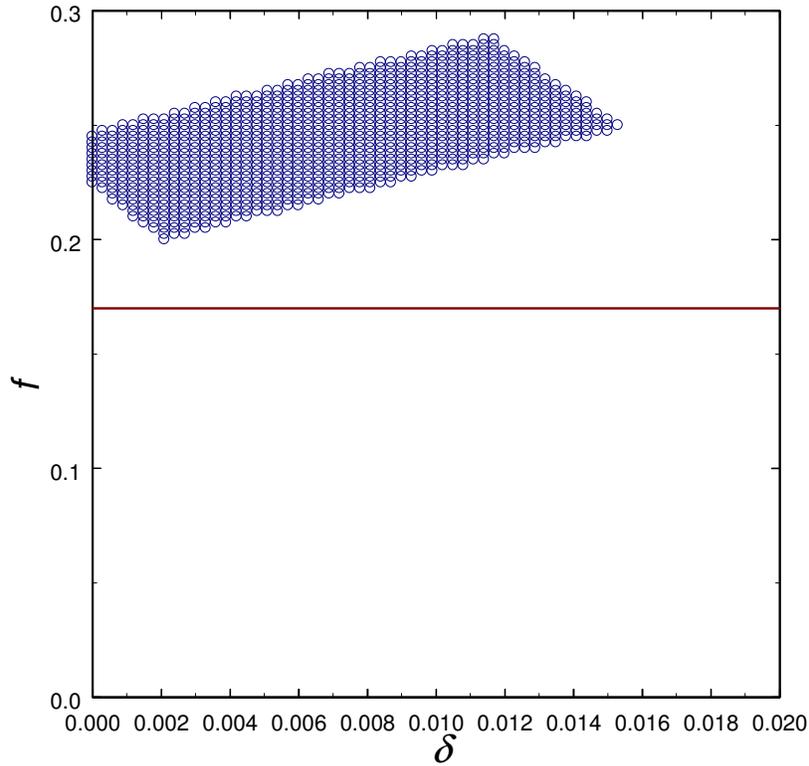

Figure 13. The region of the $\delta - f$ plane for which magnetic models of CM Dra A have evolutionary tracks which pass through the error boxes. The GTC criterion has been used for the onset of convection in the presence of magnetic field, and $\delta =$ const at all radial locations. The systematic shift in the observed radii due to spots is assumed proportional to the effective fractional spot coverage, $f$. The horizontal line is placed at $f = 0.17$ (Morales et al. 2010).

7.3.3. Modeling the onset of magneto-convection in CM Dra: the "ceiling option"

In the CM Dra models we have described up to this point, in applying the magnetic inhibition of convective onset, we have reported results only for the limiting case with $\delta =$ const at all radial locations (i.e. option (i) in Section 5.2 above). Now we consider the "ceiling" option (iv) in Section 5.2. That is, we choose a value of $\delta$ which is fixed at a constant value from the surface in to a certain depth, or radial location: at that location, the local magnetic field strength reaches a prescribed value (the "ceiling"). At deeper radial locations, the field is held fixed at its ceiling value. In Fig. 14, we present results in the same format as those in Figs. 12 and 13, except that now the magnetic field strength is not allowed to become any larger than $10^6$ G. Results are shown in Fig. 14.

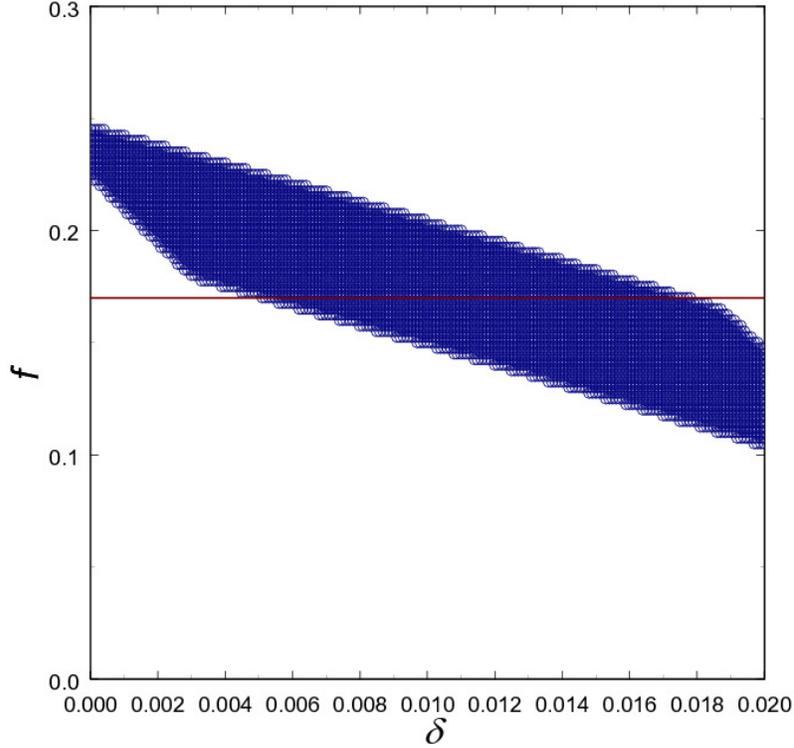

Figure 14. Magnetic models of CM Dra A in which we apply option (iv) (see Section 5.2). The abscissa refers to the value of $\delta$ in the near-surface layers. Magnetic fields inside the star are nowhere stronger than $10^6$ G. The figure shows the region of the $\delta - f$ plane for which CM Dra A tracks pass through the observed error boxes. The GT criterion has been used for the onset of convection in the presence of magnetic field. The systematic shift in the observed radii due to spots is assumed proportional to the effective fractional spot coverage, $f$. The horizontal line is placed at $f = 0.17$ (Morales et al. 2010).

As can be seen from Fig. 14, the spot coverage preferred by Morales et al. (2010), i.e. $f = 0.17$, can be replicated with a broader range of $\delta$ choices ($\delta = 0.004$-$0.018$) in our "ceiling" models than in the constant-$\delta$ models (see Figs. 12 and 13). In fact, our models can also fit the data with considerably smaller spot coverages than Morales et al prefer. Our "ceiling" models can fit the data with spot coverages as small as $f \approx 0.1$. With this coverage, our models fit the observed data provided that the near-surface values of $\delta$ are in the range $\approx 0.020$-$0.025$.

The fact that we can replicate the observed stellar parameters of CM Dra (including the spot-reduced radius) with an interior field of no more than $10^6$ G is an important change in the conclusions we derived from results in Figs. 12 and 13. In the latter models, we were required to admit the presence of fields in excess of $10^7$ G in the deep interior. Now we find that the observed data can be fit with interior fields which are an order of magnitude weaker. We shall return to this point when we discuss dynamo action in Section 8.

We note that the near-surface values of $\delta$ =0.020-0.025 which we derive for the "ceiling models" are definitely larger than those we obtained from Figs. 12 and 13. In other words, in order to fit the same data with a "ceiling model", the *surface* field strengths in the "ceiling" models in Fig. 14 must be larger than those in the constant-$\delta$ models in Figs. 12 and 13. This behavior is readily understandable: since the interior fields in Fig. 14 are some 10 times weaker than the interior fields in Figs. 12 and 13, it follows that, in order to replicate the same data set with Fig. 14, we need to offset the *reduced* magnetic effects in the *interior* with *increased* magnetic effects in the *exterior* parts of the star.

With values of $\delta$ =0.020-0.025, we ask: given the increase in gas pressure $P$ as we move beneath the surface, at what depth in the star does the magnetic field rise to its "ceiling value" of 1 MG? The answer is, at the depth where $P = 2 \times 10^{12}$ dyn cm$^{-2}$ : this occurs at $0.98R$ in the model with $\delta = 0.02$, $f = 0.1$, and also in the model with $\delta = 0.012$, $f = 0.17$.

7.3.4. Population II abundances

So far, we have considered how the presence of spots affects models of population I composition. Kinematically CM Dra appears to be a population II object, and might have a sub-solar heavy element abundance. Models with lower heavy element abundance are in general hotter and smaller than solar abundance models. As a consequence, for a population II model to fit CM Dra requires either the mixing length ratio to be lower or a larger spot coverage or a larger value of the magnetic inhibition parameter or some combination of these effects. For example, if $Z = Z_0/3$ and $\alpha$ takes the solar value, to fit CM Dra requires a spot coverage $f = 0.24 – 0.28$, depending on the helium abundance, and $\delta \sim 0.01$ for the GT magneto-convection criterion.

7.3.5. CM Dra B

In the above, we have reported our modeling results only for the A component of CM Dra. We have also applied our models to CM Dra B. However, because of the larger error in luminosity, this does not add any additional constraints on the convective efficiency parameters. However, the models indicate that we cannot exclude the conclusion that the same regions of $f$ - $\delta$ space lead to acceptable fits for both A and B components. This is consistent with the expectation that dynamos with comparable efficiency are expected in these two nearly identical stars in a synchronized system.

## 8. Comparison with observations and dynamo models of cool stars

Our models provide estimates of field strengths in two interesting locations: at the visible surface, and deep in the interior. Let us compare these estimates with other available information.

First, as regards surface field, we note that in our magneto-convection models of CM Dra A, the numerical values of $\delta$ in the best-fitting models are no larger than 0.020-0.025. These limits can be converted to an upper limit on the strength of the (vertical component of the) surface magnetic field, provided we know the gas pressure $P$ at the photosphere. The latter emerges from our model, $P(\text{photosphere}) = 5 \times 10^5$ dyn cm$^{-2}$. Inserting this along with $\delta = 0.020$-$0.025$ into eq. (3) above (with $\gamma = 5/3$), our results indicate that the vertical (poloidal) fields on the surface of CM Dra A (and CM Dra B) have strengths of 460-510 G.

Is there any empirical evidence to support these estimates? There are no direct measurements of magnetic fields in CM Dra. However, Saar (2010) has suggested that, given the spectral type and rotational period of CM Dra, the magnitude of the surface fields might be as strong as 1-2 kilogauss. Such fields would probably be representative of active regions and starspots, i.e. they might be associated with the toroidal component of the stellar field. From solar observations, we know that the toroidal fields build up their strengths by having the differential rotation interact with the poloidal field: stretching of the latter leads to the strong fields which are observed in active regions and spots. Therefore, the poloidal field strengths should provide a firm *lower* limit on the toroidal field strengths. Our finding of several hundred gauss for the surface value of (poloidal field) $B_v$ is consistent with the stronger value (of toroidal field) estimated by Saar.

Moreover, we note that surface fields of several hundred G are consistent with values reported by observers of other early- and mid-M dwarfs (Reiners & Basri 2009).

The (vertical) field strengths we derive for the surface of CM Dra are up to 50-100 times stronger than the poloidal fields in the Sun (5-10 G: cf. MMT07). This is presumably related to the 22-times shorter rotation-period of CM Dra. Since flare energies scale as $B^2$, the presence of stronger fields in CM Dra correspond to flare energies in CM Dra which could exceed solar flare energies by some orders of magnitude. This is consistent with observational evidence summarized in Section 1.3 above.

Turning now to the (unobservable) fields in the interior, the only comparison which is available is with another modeling effort. Browning (2008) has reported 3-D large-eddy compressible MHD simulations of dynamo action in fully convective stars with mass 0.3 $M_\odot$, i.e. with a mass close to that of CM Dra A. The small-scale effects are modeled with certain choices of numerical values for eddy diffusivities (viscosity, thermal, and magnetic): because of lack of information, Browning chooses to keep these diffusivities constant with radius (akin to our option (i) in Section 5.2). In all 6 models reported by Browning, the angular velocity is fixed at the solar value. The models achieve dynamo action, with amplification of seed fields to (rough) equipartition between magnetic energy and convective kinetic energy. The large-scale mean fields in the interior of the models reach maximum values of 13 kiloGauss. Browning points out that the magnitudes of these fields in the simulations are "likely sensitive at some level to the choices of the dimensionless numbers Pm (magnetic Prandtl number) and Rm (magnetic Reynolds number)": the physical values of these dimensionless numbers in a "real" star are orders of magnitude different from those used in the simulations. It is not yet known what changes might occur in the simulations if different values were chosen for the dimensionless numbers, and/or if different choices were made for the boundary conditions (B.C.'s) (there are 12 B.C.'s to be specified in the code). It would not be surprising if, when all aspects of parameter space are explored, the maximum fields could differ from those reported by Browning (in the small number of simulations which have been run thus far) by factors of a few.

Before we can compare our results with those of Browning, we note that Browning's calculations were made for a single value of the rotational angular velocity $\Omega = 2.6 \times 10^{-6}$ sec$^{-1}$: this value is appropriate for a convective star which rotates in a period of 28 days, such as the Sun. However, in order to compare more realistically with CM Dra A, with its rotational period of 1.27 days, simulations would have to be done with $\Omega$ values which are larger than Browning used by a factor of 22. How might the dynamo field strength $B$ in a star change as $\Omega$ increases? Intuitively, the expectation is that increasing $\Omega$ should lead to increasing $B$. This is based, at least qualitatively, on two empirical correlations which are known to exist among cool stars. First, there is a well-established connection between rotation and chromospheric/coronal activity (Pallavicini et al. 1981): faster rotation is correlated with higher levels of activity, and (at least in the Sun) activity is observed to be associated with release of magnetic energy. Second, magnetic moments of cool stars have been found to be positively correlated with angular momenta (Arge et al. 1995).

However, neither of these empirical correlations provides a reliable scaling between the particular variables $B$ and $\Omega$: there are too many other factors involved in the empirical correlations to allow us to isolate the effects of $B$ and $\Omega$ alone. We need instead to look to dynamo theory. Kippenhahn (1973) cites two possible scalings for a stellar dynamo: $B \sim \Omega$ and $B \sim \Omega^{3/2}$. Both of these scalings are at the very least consistent with the intuition that as $\Omega$ increases, $B$ should increase also. Moreover, we have already pointed out that our estimates of (poloidal) fields $B_v$ in CM Dra are 50-100 times stronger than those in the Sun: since $\Omega$ is 22 times larger in CM Dra than in the Sun, a power law-scaling between these values of $B_v$ and $\Omega$, i.e. $B_v \sim \Omega^\varepsilon$, would be consistent with our estimates if $\varepsilon \approx 1.27$-$1.49$. This range of power law indices is entirely consistent with the two scalings cited by Kippenhahn (1973).

To the extent that Kippenhahn's scalings apply to stellar simulations, we find that an increase by 22 in $\Omega$ should lead to values of $B$ which are larger than those of Browning by factors of 22-103. Rather than maximum fields of 13 kG, as Browning reports in a star rotating with the solar angular velocity, the maximum fields in the fast rotating CM Dra could be as large as (0.3-1.3) megaGauss (MG). And if the maximum $B$ values of Browning's simulations are indeed subject to errors of a few, the lower limit of this range in CM Dra could approach 1 MG.

Fields of order 1 MG are in agreement with the values we have imposed in our "ceiling" models (Section 7.3.3) which provide the best fit to the empirical data (including spot-corrected radii) in CM Dra.

## 9. Discussion and Conclusions

CM Dra is an interesting system: it contains the lowest mass main sequence stars for which precise masses and radii have been determined. In the present paper, we have found it a challenging exercise to use a stellar evolution code to fit the empirical *L* and *R* values of the components of CM Dra which were reported by Morales et al (2009). Our goal has been to derive stellar models with parameters which lie within the statistical error bars, and simultaneously agree with the reported age of 4.1 Gyr. The challenge arises in part from the unprecedentedly small statistical errors reported in *L* and *R*.

After searching through the (non-magnetic) parameter space of *Y*, *Z*, *α* (the mixing length parameter), and opacity, we were not able to find any consistent non-magnetic model of CM Dra, even if we extended the search to rather implausible values of *Y* and *Z*.

In order to address this inconsistency, we note that there is strong observational evidence for magnetic fields in CM Dra. In the presence of such fields, the onset of convection is inhibited, and this effect can be modeled (see MM01) in terms of an inhibition parameter *δ* (eq. (3)) which is proportional to the square of magnetic field strength. But even when we include magnetic inhibition of convection, we are not successful in fitting the parameters of CM Dra as reported by Morales et al (2009) within the statistical errors, unless the age of CM Dra is no more than 0.3 Gyr. Such a young age seems unlikely in view of the high space motion, and also in view of the age (4.1 Gyr) of the white dwarf which appears to be a common proper motion companion of CM Dra.

One of the pieces of evidence for the occurrence of magnetic fields in CM Dra is the presence of sinusoidal variations in brightness of the stars *outside* eclipses, with amplitudes at the 1-2% level: these variations are interpreted to mean that there are starspots on the surface. The shape of the light curves (specifically, the absence of flat portions of the light curves) indicates that some of the starspots are so close to the poles of the star that they undergo only minimal rotational modulation. Morales et al. (2010) show that the presence of polar spots leads to a systematic overestimate in extracting the value of the stellar radius from the analysis of eclipses. The magnitude of the overestimate in $R$ depends on the numerical value of the effective spot coverage, $f$. For values of $f$ suggested by Morales et al. (2010) in CM Dra, i.e. $f \approx 0.17$, the stellar radii need to be reduced by a few percent.

These spot-related reductions in radius are larger than the statistical errors in the radius which were reported by Morales et al (2009). When the spot-corrected values of radius are used, we find that, in the context of reduced $\alpha$, we can identify regions of $f$-$\alpha$ space which allow fits to the CM Dra data. Unfortunately, reductions in $\alpha$ are not readily converted into estimates of magnetic field strengths.

On the other hand, in the context of the magnetic inhibition parameter $\delta$, there is a direct connection between $\delta$ and the local magnetic field strength. We have successfully identified regions of $f$ - $\delta$ space which allow us to fit the CM Dra data. We can fit the data by fixing $\delta$ at a constant value in the outermost layers of the star, but then limiting the field strength inside the star to no more than $10^6$ G. We argue that interior fields of such a strength ($10^6$ G) may be quite consistent with rotationally-enhanced dynamo simulations of convective stars. As regards conditions at the surface of the star, our values of $\delta$ can be directly converted to field strength at the surface of CM Dra. The numerical values which we derive for $\delta$ in CM Dra A are at most of order 0.020-0.025: using the photospheric pressures in our models, we find that these values of δ correspond to (vertical) fields of about 500 gauss on the surface of the star.

Our model fits suggest that comparable field strengths are present on CM Dra A and B. This is consistent with synchronized rotation of both components in this tidally locked system.

Finally, we note that, for fully convective stars, we find that the effects on the stellar structure of adding cool spots are equivalent to reducing the mixing length ratio.

**Acknowledgments**

This work has been supported in part by the NASA Delaware Space Grant program and by a grant from the Mount Cuba Astronomical Foundation. DJM thanks Steve Saar for estimates of the surface fields on CM Dra, and Yilen Gomez for a helpful discussion..

**References**

Arge, C. N., Mullan, D. J., & Dolginov, A. Z. 1995, ApJ 443, 795
Browning, M. K. 2008, ApJ 676, 1262
Chabrier, G. & Baraffe, I. 1995, ApJ Lett 451, L29

Chabrier, G., Gallardo, J., & Baraffe, I. 2007, A&A 472, L17

Chandrasekhar, S. 1961, *Hydrodynamic and Hydromagnetic Stability* (Oxford: Clarendon press), p. 189

Christensen-Dalsgaard, J. et al. 1996, Science, 272, 1286

Cox, A. N., Shaviv, G., & Hodson, S. W. 1981, ApJL 245, L37

Deeg, H. et al. 1998, A&A 338, 479

Eggen, O. J, & Sandage, A. 1967, ApJ 148, 911

Garcia, R. A., Mathur, S., Salabert, D., Ballot, J., Regulo, C., Metcalfe, T. S. & Baglin, A. 2010, Science 329, 1032

Gershberg, R. E. 2002, Solar Type Activity in main Sequence Stars (in Russian), (Odessa: Astroprint), p. 312

Gough, D. O. & Tayler, R. J. 1966, MNRAS 133, 85 (GT)

Gudel, M., Telleschi, A., Skinner, S. L., Audard, M., & Ness, J.-U. 2005, in Proc. 13th Cool Stars Workshop, eds. Favata et al., ESA SP-560, p. 605

Jenkins, J. S., Ramsey, L. W., Jones, H. R. A., Pavlenko, Y., Gallardo, J., Barnes, J. R. & Pinfield, D. J. 2009, ApJ 704, 975

Kim, S.-L., Chun, M.-Y., Lee, W.-B., & Doyle, L. 1997, Inf. Bull. Var. Stars number 4462

Kippenhahn, R. 1973, in Stellar Chromospheres IAU Colloq. 19, ed. S. Jordan and E. Avrett, NASA SP-317, p. 265.

Kozhevnikova, A. V., Svechnikov, M. A., & Kozhevnikov, V. P. 2009, Astrophysics, 52, 512

Lacy, C. H. 1977, ApJ 218, 444

MacDonald, J. & Mullan, D. J. 2009, ApJ 700, 387 (MM09)

Metcalfe, T. S., Mathieu, R., Latham, D. W., & Torres, G., 1996, ApJ, 456, 356

Morales, J. C. et al. 2009, ApJ 691, 1400

Morales, J. C., Gallardo, J., Ribas, I., Jordi, C., Baraffe, I., & Chabrier, G. 2010, ApJ 718, 502

Mullan, D. J. 1974, ApJ 192, 149

Mullan, D. J. 2010, ApJ 721, 1034

Mullan, D. J. & MacDonald, J. 2001, ApJ 559, 353 (MM01)

Mullan, D. J. & MacDonald, J. 2010, ApJ 713, 1249 (MM10)

Mullan, D. J., MacDonald, J., & Townsend, R. H. D. 2007, ApJ 670, 1420 (MMT07)

Nelson, T. E. & Caton, D. B. 2007, Inf. Bull. Var. Stars number 5789.

Olive, K. A., Skillman, E. D. 2004, ApJ, 617, 29

Paczyński, B. & Sienkiewicz, R. 1984, ApJ 286, 332

Pallavicini, R., Golub, L. Rosner, R., Vaiana, G. S., Ayres, T., & Linsky, J. L. 1981, ApJ 248, 279

Reiners, A. & Basri, G. 2009, A&A, 496, 787

Saar, S. H. 2010, personal communication to DJM.

Saar, S. H. & Bookbinder, J. A., 1998, in R. A. Donahue and J. A. Bookbinder (eds.), ASP Conf. Series Vol. 154, p. 2042

Spruit, H. C. 1992, in Surface Inhomogeneities on Late-Type Stars, Proceedings of a colloquium held at Armagh Observatory, Northern Ireland, 24-27 July, 1990. Edited by P. B. Byrne and D. J. Mullan. Lecture Notes in Physics, Vol. 397. Published by Springer-Verlag, Heidelberg, Germany, 1992., p.78

Spruit, H.C. & Weiss, A. 1986, A&A, 166, 167

Torres, G., Andersen, J., & Gimenez, A. 2010, Astron. Astrophys. Rev. 18, 67

Van de Kamp, P. and Lippincott, S. L. 1951, PASP 63, 141

Vilhu, O., Ambruster, C. W., Neff, J. E., Linsky, J. L., Brandenburg, A., Ilyin, I. V., Shakhovskaya, N. I. 1989, A&A 222, 179

Zahn, J.-P., 1977, A&A 57, 383